\documentclass[10pt,conference]{IEEEtran}
\IEEEoverridecommandlockouts
\PassOptionsToPackage{hyphens}{url}\usepackage{hyperref}
\usepackage{cite}
\usepackage{amsmath,amssymb,amsfonts}
\usepackage{algorithm}
\usepackage{algpseudocode}
\algtext*{EndFor}
\algtext*{EndFunction}
\usepackage{graphicx}
\usepackage{textcomp}
\usepackage{xcolor}
\usepackage{subfig}
\usepackage[percent]{overpic}

\usepackage{todonotes}

\usepackage{physics}
\DeclareMathAlphabet{\mathpzc}{OT1}{pzc}{m}{it}

\def\BibTeX{{\rm B\kern-.05em{\sc i\kern-.025em b}\kern-.08em
    T\kern-.1667em\lower.7ex\hbox{E}\kern-.125emX}}
\begin{document}

\title{Cross Entropy Hyperparameter Optimization for Constrained Problem Hamiltonians Applied to QAOA}

\author{\IEEEauthorblockN{Christoph Roch, Alexander Impertro, Thomy Phan, Thomas Gabor, Sebastian Feld and Claudia Linnhoff-Popien}
\IEEEauthorblockA{\textit{LMU Munich} \\
Germany\\
christoph.roch@ifi.lmu.de}
}

\maketitle

\begin{abstract}
Hybrid quantum-classical algorithms such as the Quantum Approximate Optimization Algorithm (QAOA) are considered as one of the most encouraging approaches for taking advantage of near-term quantum computers in practical applications. Such algorithms are usually implemented in a variational form, combining a classical optimization method with a quantum machine to find good solutions to an optimization problem. The solution quality of QAOA depends to a high degree on the parameters chosen by the classical optimizer at each iteration. However, the solution landscape of those parameters is highly multi-dimensional and contains many low-quality local optima.
In this study we apply a Cross-Entropy method to shape this landscape, which allows the classical optimizer to find better parameter more easily and hence results in an improved performance.
We empirically demonstrate that this approach can reach a significant better solution quality for the Knapsack Problem.
\end{abstract}
\vspace{+1em}

\section{Introduction}
Over the last years more and more quantum computing companies have made their devices available for researchers \cite{IBMLaunch, DwaveLaunch}. This effort resulted in a lot of research on quantum algorithms, especially on how to develop noise-resilient algorithms, which can be executed on near-term quantum computers. For a certain number of computational problems it is known that a gate-based quantum computer can outperform any known classical algorithm \cite{grover1996fast, shor1999polynomial}. However, these algorithms often require thousands of gates in practice \cite{roetteler2017quantum}, something that is impossible to do accurately without error correction on noisy intermediate scale quantum (NISQ) devices, which currently have less than 100 qubits and high error rates. That is why many near-term algorithms have been introduced in an attempt to leverage NISQ devices. Most popular are hybrid quantum-classical algorithms, including the Variational Quantum Eigensolver (VQE) \cite{peruzzo2014variational} and the Quantum Approximate Optimization Algorithm (QAOA) \cite{farhi2014quantum}. 

These algorithms combine a gate-model based quantum computer with a classical optimizer. On the quantum side, the initial state of the system is brought into an equal superposition of basis states, and afterwards evolved by applying the algorithm specific gates, with the goal of preparing the state which minimizes the cost. This so called quantum evolution is characterized by a shallow depth circuit also commonly referred to as the ansatz, which enables the algorithm to be run on NISQ devices.

One can achieve shallow circuit depth by parameterizing the gates. Since the action of a quantum gate can be expressed as a rotation in the Hilbert space that contains the quantum state, one can consider the rotation angle as the parameter of the gate. The optimal quantum evolution can then be found by varying the parameters of a set of gates. In the variational setting of QAOA and VQE, the parameters of the gates are usually found with the help of a classical optimizer. 

On the one hand, the performance of such hybrid algorithms depend on the classical optimization method, and on the other hand on the series of parameterized gates. However regarding the solution quality, one major hurdle lies in the difficulty to efficiently optimize in the non-convex, high-dimensional landscape of the gate parameters, which contains many low-quality, non-degenerate local optima \cite{zhou2018quantum}.

In this work we address this problem by applying a Cross-Entropy (CE) method to optimize the hyperparameters of the solution landscape, which are represented by the penalty values of the constrained Hamiltonian of the optimization problem. In our case we refer to the Knapsack Problem (KP). We can show that by optimizing those penalty values the classical optimizer is able to find better gate parameters for the quantum circuit. This results in a significant increase in solution qualities as compared to the conventional approach without CE optimization.

The paper is structured as follows. Section 2 introduces the mathematical formulation of the Knapsack Problem, the corresponding Ising Model and the transformation to the operators of QAOA. Additionally, the basic principles of CE are reviewed. Section 3 gives an overview of related work regarding the Knapsack Problem and QAOA. Afterwards the concept of our QAOA-CE approach is given in Section 4 before we evaluate and discuss the results in Section 5. Finally, we conclude this paper in Section 6.

\section{Background}
\subsection{Knapsack Problem}
The Knapsack Problem is a constrained combinatorial optimization problem that refers to the general problem of packing a knapsack with the most valuable items without exceeding its weight limit. The best known variant of the KP is the 0-1 Knapsack Problem, which means that each item may be used or packed only once.

In the 0-1 Knapsack Problem, $n$ items are given, each having a certain weight $w_\alpha$ and a certain value $c_\alpha$. The items must be picked in a way that the total weight of the items is less than or equal to the knapsack capacity $W$, Eq. \eqref{eq:constraint_knapsack}, and the total sum of the item values is maximized, Eq. \eqref{eq:objective_function_knapsack}. Variable $x_\alpha$ is set 1 if the item is packed in the knapsack and 0 otherwise \cite{martello1990knapsack}.

\begin{equation}\label{eq:objective_function_knapsack}
max \sum_{\alpha = 1}^{n} c_\alpha x_\alpha 
\end{equation} 

s.t.
\begin{equation}\label{eq:constraint_knapsack}
\sum \limits_{\alpha=1}^n w_\alpha x_\alpha \le W
\end{equation}

As shown in \cite{karp1972reducibility}, this problem is NP-complete.
In this work we refer to the 0-1 KP and its term is used interchangeably with KP.

\subsection{Ising Model}
The Ising model is a theoretical mathematical model used to describe phase transitions and certain properties of particles in a physical system that evolves over time. 

Given are $n$ particles, which are attached to the vertices of a graph $G = (V, E)$. Each of these particles can be in one of two possible spin states, which have the values $-1$ and $+1$. A spin configuration $s = s_1,...,s_n$ is an assignment of the spin values of all particles. Furthermore, there are external forces $h_i$ acting on the respective particles, as well as forces $J_{ij}$ acting between neighbouring particles. The energy of a certain spin configuration is then given by \cite{mcgeoch2014adiabatic}:
\begin{equation}\label{eq:ising_model}
H(s) = \sum_i^n h_i s_i + \sum_i^n \sum_{j>i}^n J_{ij}s_i s_j.
\end{equation}

From Eq. \eqref{eq:ising_model}, it follows that the energy of the configuration, represented by a so called Hamiltonian $H$, corresponds to the sum of all vertices and the interacting forces on the edges of the graph. Many optimization problems, including the KP, can be formulated as an Ising model. Finding the solution with minimum cost then corresponds to the spin configuration with the lowest energy, i.e. the minimum of $H(s_1,...,s_n)$. 

\subsection{Knapsack Problem to Ising Model}\label{subsec:kp-to-ising}

In order to implement the KP on a quantum computer using QAOA, we need to encode the objective function of the KP into a Hamiltonian which is diagonal in the computational basis.

The weight constraint \eqref{eq:constraint_knapsack} can be encoded in the following quadratic Hamiltonian, as stated in \cite{lucas2014ising}:
\begin{equation}\label{eq:constraint_knapsack_ising}
H_1 = A \left( 1 - \sum_{n=1}^{W} y_n \right)^2 + A \left( \sum_{n=1}^{W} n y_n - \sum_{\alpha = 1}^{N} w_\alpha x_\alpha \right) ^2
\end{equation}
while the objective function in \eqref{eq:objective_function_knapsack} is straightforwardly
\begin{equation}\label{eq:objective_function_knapsack_ising}
H_2 = - B \sum_{\alpha = 1}^{N} c_\alpha x_\alpha.
\end{equation}

Here, $y_n$ for $1 \leq  n \leq W$ is a binary variable, which is set 1 if the final weight of the knapsack is $n$ and 0 otherwise. $H_1$ enforces that the weight can only take exactly one value and that the weight of the items in the knapsack equals the value we claimed it did. The parameters $A,B$ are chosen according to $0 < B \cdot \text{max}(c_\alpha) < A$ in order to penalize violations of the weight constraint (i.e., adding one item to the knapsack, which makes it too heavy, is not allowed). It is mentionable, that one can reduce the number of binary variables using the so called log trick to $N + \lfloor 1 + log W \rfloor$ \cite{lucas2014ising}.

To implement the problem Hamiltonian $H_3=H_1 + H_2$ on a gate model quantum computer, we need to express it as a linear combination of one- and two-spin variables. Using the transformations $x_\alpha=\frac{1}{2}(s_\alpha + 1)$ and $y_n=\frac{1}{2}(u_n + 1)$, we obtain for the weight Hamiltonian of Eq. \eqref{eq:constraint_knapsack_ising} the following Ising Hamiltonian:
\begin{equation}\label{eq:constraint_knapsack_ising_ex}
\begin{gathered}
H_{1^*} = A \Bigg\{ \sum_n \left[ \frac{W}{2} - 1 + n\left(\frac{W^2+W}{4} - \frac{1}{2} \sum_{\alpha} w_\alpha \right) \right] u_n \\+ \frac{1}{2}\sum_{n<l} (1+nl) u_n u_l + \frac{1}{2} \sum_{\alpha}\left[ -\frac{W^2+W}{2} + \sum_{\beta} w_\beta \right] w_\alpha s_\alpha \\+ \frac{1}{2}\sum_{\alpha < \beta} w_\alpha w_\beta s_\alpha s_\beta - \frac{1}{2}\sum_{n, \alpha} n w_\alpha u_n s_\alpha \Bigg\}
\end{gathered}
\end{equation}
Similarly, the cost Hamiltonian of Eq. \eqref{eq:objective_function_knapsack_ising} becomes:
\begin{equation}\label{eq:objective_function_knapsack_ising_ex}
H_{2^*} = -B\sum_{\alpha}\frac{c_\alpha}{2} s_\alpha
\end{equation}
In both derivations, constant energy offsets were omitted. Finally, the complete problem Hamiltonian $H_{3^*}$ is given as the sum of \eqref{eq:constraint_knapsack_ising_ex} and \eqref{eq:objective_function_knapsack_ising_ex}: 
\begin{equation}\label{eq:objective_function_knapsack_ising_ex_total}
H_{3^*} = H_{1^*} + H_{2^*}
\end{equation}

\subsection{Quantum Approximate Optimization Algorithm}
The Quantum Approximate Optimization Algorithm \cite{farhi2014quantum} is a hybrid quantum-classical algorithm specifically developed for approximately solving combinatorial optimization problems on a gate model quantum computer.

In order to manipulate states on a gate model quantum computer, the Ising Hamiltonian of our optimization problem, as defined in Eq. \eqref{eq:objective_function_knapsack_ising_ex_total}, must be described by an operator. Since the measurement results of the operator (eigenvalues) correspond to the cost of the optimization problem, the Pauli Z-operator $\sigma^Z$ with \begin{equation*}\sigma^Z = \begin{pmatrix} 1 & 0 \\ 0 & -1  \end{pmatrix}\end{equation*} is used, whose eigenvalues ($\pm1$) correspond to the positive and negative spin values of the Ising Hamiltonian. By replacing the spin variables $s_{\alpha}, s_{\beta}, u_i$ and $u_j$ with $\sigma^Z$ and each term of the form $s_\alpha s_\beta, u_i u_j, u_i s_\alpha $ by $\sigma^Z \otimes \sigma^Z$, the desired cost Hamiltonian $H_C$ is obtained.

The variational form of QAOA first starts from a uniform superposition state of every possible solution. 
A trial state is prepared by evolving the system under the cost and an additional driver Hamiltonian, which is used to explore the solution landscape. One possible implementation for the driver Hamiltonian is given by \begin{equation*}\label{eq:driver}
H_D = \frac{1}{2} \sum_{i} \sigma^X_i,
\end{equation*}
where $\sigma^X$ is the Pauli X-operator. Since this evolution is usually difficult to implement, it is approximated using the Trotter-Suzuki expansion \cite{suzuki1976generalized}. This approximation is achieved by repeated applications of the trotterized cost and driver propagator,
\begin{equation}\label{eq:evolve}
\ket{\beta,\gamma} = V_pU_p ... V_2U_2V_1U_1\ket{\psi},
\end{equation}
where ${U_p = exp(-i\gamma_pH_C)}$ is the cost propagator and ${V_p = exp(-i\beta_pH_D)}$ the driver propagator. The cost Hamiltonian $H_C$ is evolved for some time $\gamma_p$, while the driver Hamiltonian $H_D$ is evolved for some time $\beta_p$. The length $p$ of the alternating sequence in Eq. \eqref{eq:evolve} determines the degree of approximation. 

Within QAOA the variational parameters $(\beta,\gamma)$ of the gates, as stated in Eq. \eqref{eq:evolve}, are used to prepare the trial state $\ket{\psi (\beta,\gamma)}$ on the quantum processor by applying the alternating sequence of propagators. The state is then measured and the result is used by a classical optimizer to find new parameters $(\beta,\gamma)$, with the goal of finding the ground-state energy of the cost Hamiltonian $min \bra{\psi (\beta,\gamma)} H_C \ket{\psi (\beta,\gamma)}$. The ground state corresponds to the global optimum of the classical optimization problem. This iterative process continues until the classical optimizer converges or a solution of acceptable quality is found \cite{shaydulin2019multistart}.

\subsection{Cross-Entropy Method}

The Cross-Entropy method is a Monte Carlo method for importance sampling and optimization, and is known to perform well on combinatorial optimization problem with noisy objective functions \cite{rubinstein2004cross, rubinstein1999cross}.

Regarding our approach we align ourselves to a common CE algorithm, see Alg. \ref{algorithm:cross-entropy}, as stated in \cite{weinstein2013open}.
The iterative process consists of sampling a set of points $a_1...a_n$ from the distribution $\mathpzc{p}$, based on its current parameterization $\Phi_{g-1}$ (line 3). The objective function $f$ of the optimization problem assigns values $v_1...v_n$ to each point $a_1...a_n$ (line 4). After that, a selection routine picks the fraction $\rho$ of elite samples (line 5 and 6) and the computation of the new parameterization of $\mathpzc{p}$, $\Phi_{g}$ used in the next iteration based on the elite samples, is done (line 6).

\begin{algorithm}
	\caption{Cross-Entropy}\label{algorithm:cross-entropy}
	\begin{algorithmic}
		\Function{OPTIMIZE}{$\mathpzc{p},\Phi_{0},f,\rho,n,G$}
		\For{$g = 1 \to G$}
		\State $a_1...a_n \sim \mathpzc{p}(\cdot|\Phi_{g-1}), \textbf{a}\gets a_1...a_n$
		\State $v_1...v_n \sim f(a_1)...f(a_n), \textbf{v}\gets v_1...v_n$
		\State $\text{sort \textbf{a} according to \textbf{v}}$
		\State $\Phi_{g} \gets \text{argmax}_{\Phi}\prod_{i=1}^{\lceil n\rho\rceil} \mathpzc{p}(a_i|\Phi)$
		\EndFor
		\Return $a_1$
		\EndFunction
	\end{algorithmic}
\end{algorithm}

A central parameter for the algorithm is the distribution $\mathpzc{p}$, since it determines the choice of new sample points \textbf{a} for each generation. The closer the initial distribution $\mathpzc{p}(\cdot|\Phi_0)$ reproduces optimal samples, the faster the algorithm converges. When however no information about the optimal distribution is available, or it is inefficient to sample from, it is generally a good idea to choose a distribution which covers the entire sample space. This increases the probability for the algorithm to evolve towards a good solution already in early generations. The parameterization $\Phi_{g}$ of the distribution is updated according to a maximum likelihood estimate of the chosen elite fraction in the current generation. For a multivariate Gaussian distribution, the update rule is:

\begin{equation*}\label{eq:mu}
\mu_g = \frac{\sum_{i=1}^{\lceil n\rho\rceil} X_i}{\lceil n\rho\rceil} 
\end{equation*}
\begin{equation*}\label{eq:sigma}
\sigma_g^2 = \frac{\sum_{i=1}^{\lceil n\rho\rceil} (X_i - \mu_g)^T(X_i-\mu_g)}{\lceil n\rho\rceil} 
\end{equation*}
\begin{equation*}\label{eq:param}
\Phi_{g} = \langle \mu_g, \sigma_g^2 \rangle.
\end{equation*}

In addition to the choice of sampling distribution, especially the selected fraction $\rho$ of elite samples, the population size $n$ and the number of generations $G$ must be carefully adjusted for a given problem in order to maximize the likelihood of finding a good solution. In section \ref{sec:qaoa-ce}, we explain how this CE method is adapted for our specific problem.

\section{Related Work}
The general QAOA was introduced by Farhi et al. in 2014 \cite{farhi2014quantum}. They showed that QAOA is able to outperform the classical naive algorithm on the Max Cut optimization problem on 3-regular graphs and QAOA circuit length $p=1$. However, an improved classical algorithm was introduced afterwards that outperformed QAOA on this problem \cite{barak2015beating}. Nevertheless, since then a lot of research regarding QAOA and the Max Cut problem has been done both empirically as well as theoretically. Crooks \cite{crooks2018performance}, for example shows that QAOA can exceed the performance of the Goemans-Williamson \cite{goemans1995improved} algorithm for Max Cut, while Zhou et al. theoretically investigate the performance of QAOA \cite{zhou2018quantum}.

Additional theoretical work deals with the question of how many qubits are required for meaningful quantum speedups \cite{guerreschi2019qaoa, preskill2018quantum}. 

Hadfield et al. present a modification of QAOA and provide a compendium-style application of their approach to several optimization problems, including, among others, constrained problems like the Traveling Salesman Problem or Min Clique Cover Problem \cite{hadfield2019quantum}.
To our knowledge there is only one investigation regarding QAOA and the Knapsack Problem \cite{de2019knapsack}.
The authors present two Knapsack QAOA variants. Both approaches investigate how to tackle constrained problems, like the KP, with QAOA. The results reveal, that an approach with a quadratic penalty may not work properly with shallow depth circuits. They suspect the constraint to be weighted too heavily in the objective function compared to the optimization part of the Hamiltonian. 

In this work we further investigate this problem by introducing CE to QAOA to optimize the penalty values of the Knapsack problem Hamiltonian and thus improve the overall performance of QAOA.

\section{QAOA with Cross-Entropy}\label{sec:qaoa-ce}

\begin{figure*}[ht]
	\subfloat[\textit{Solution landscape (A=3.2,B=0.2)}\label{fig:landscape_bad}]{%
		\includegraphics[width=0.46\textwidth]{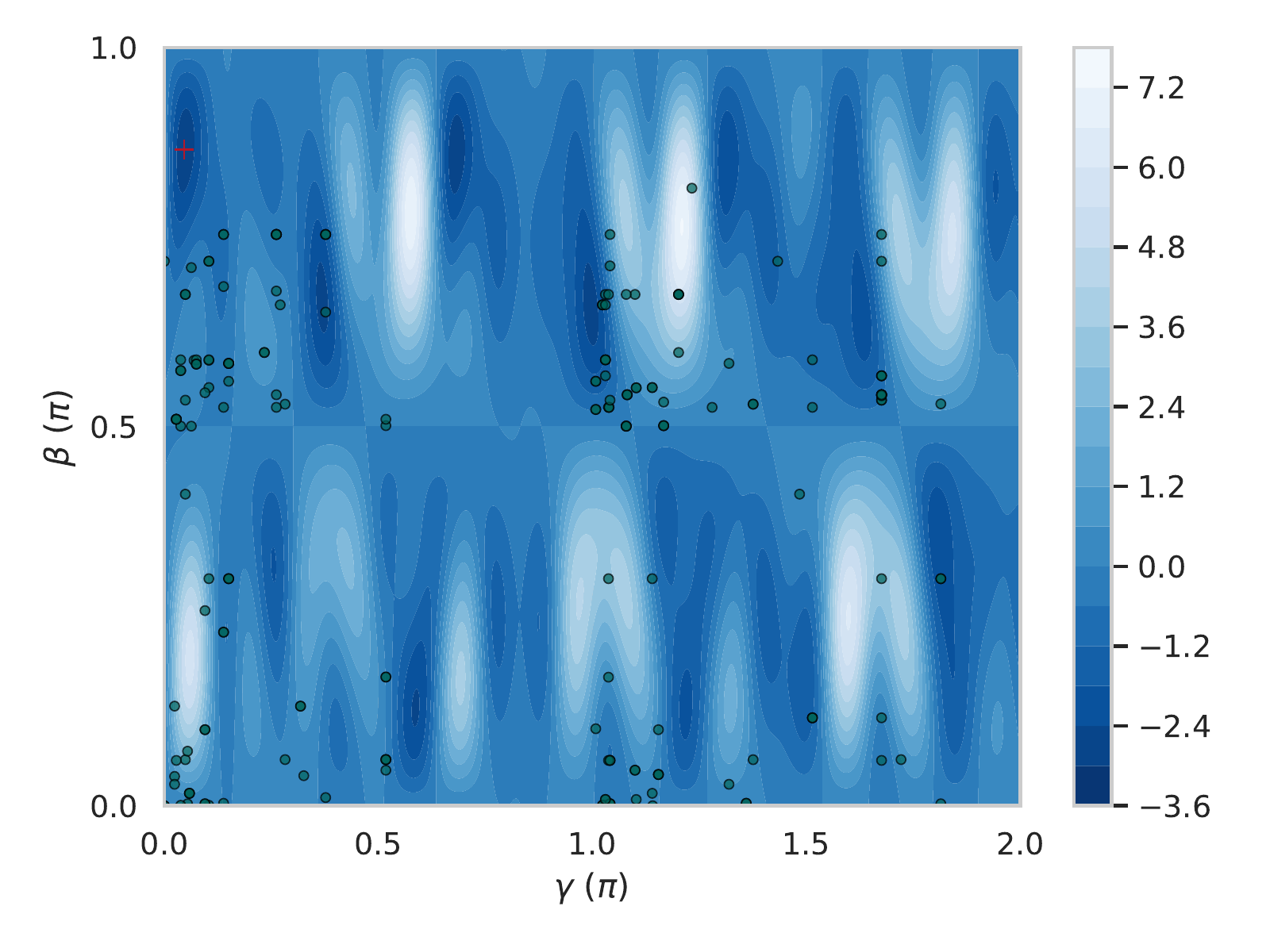}
	}
	\hfill
	\subfloat[\textit{Solution landscape (A=2.7,B=1.1)}\label{fig:landscape_good}]{%
		\includegraphics[width=0.46\textwidth]{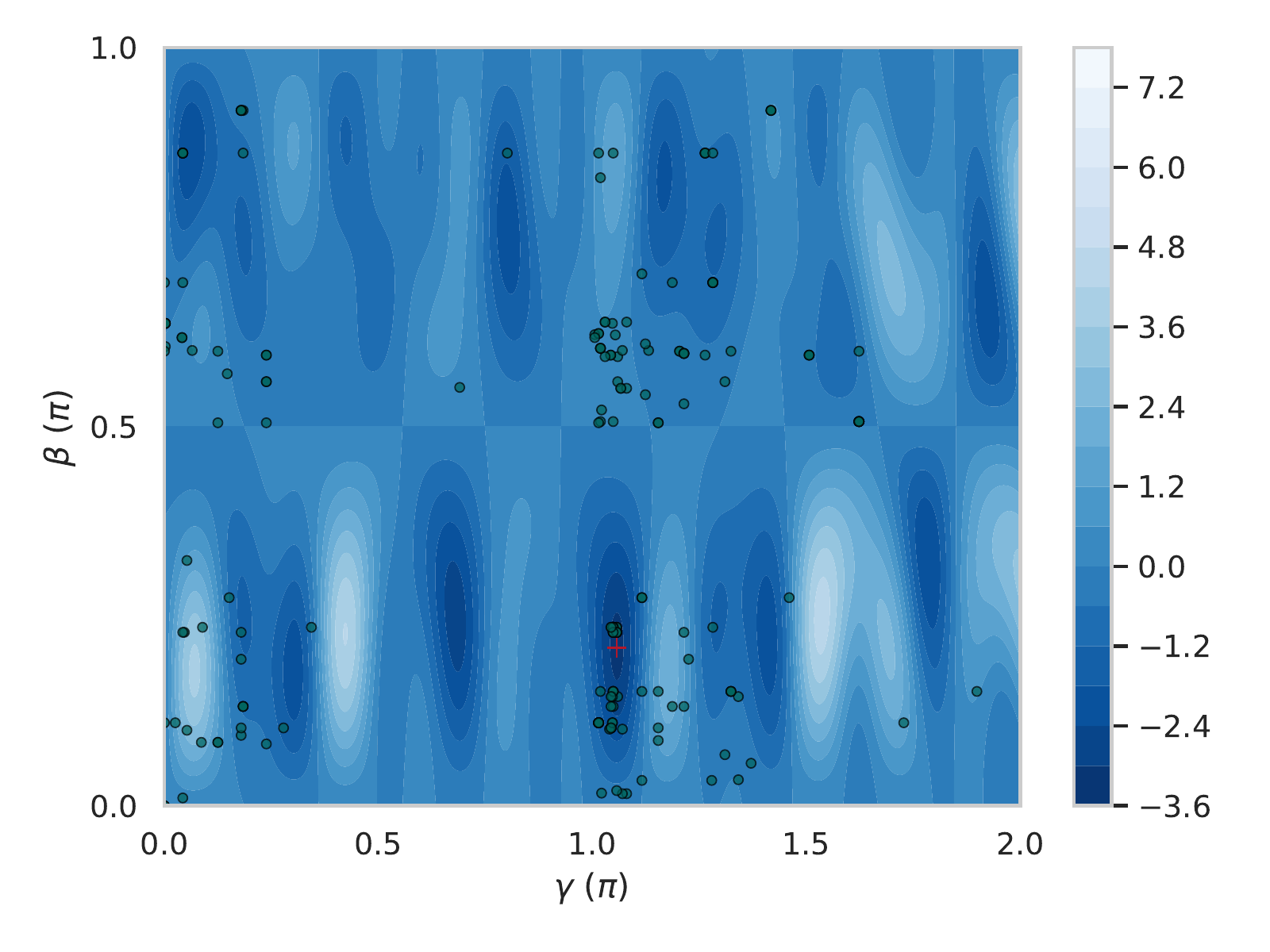}
	}
	\\
	\subfloat[\textit{Solution Quality (A=3.2,B=0.2)}\label{fig:histogram_bad}]{%
		\includegraphics[width=0.46\textwidth]{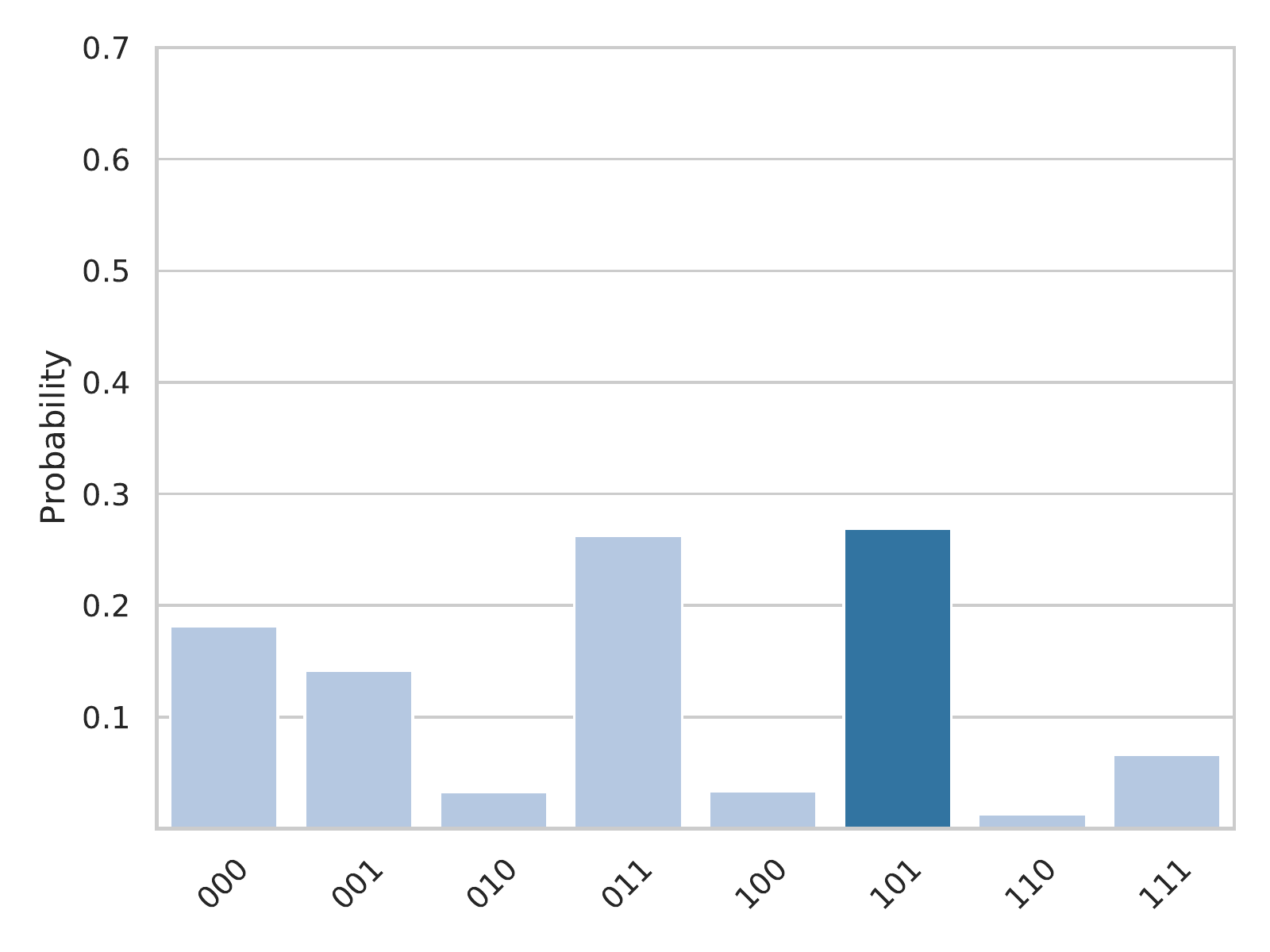}
	}
	\hfill
	\subfloat[\textit{Solution Quality (A=2.7,B=1.1)}\label{fig:histogram_good}]{%
		\includegraphics[width=0.46\textwidth]{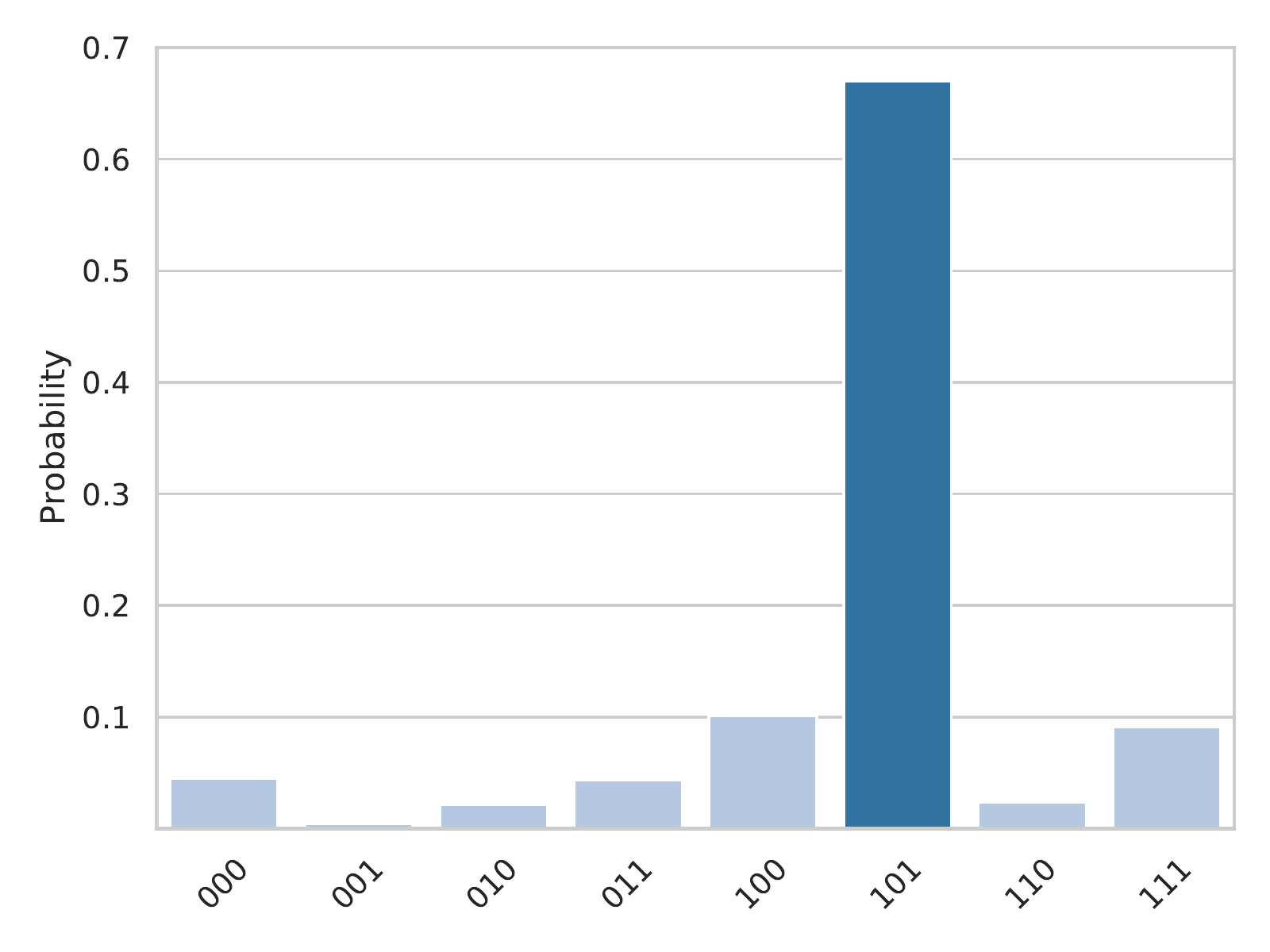}
	}
	\caption{Energy landscapes of QAOA objective function $min \bra{\psi (\beta,\gamma)} H_C \ket{\psi (\beta,\gamma)}$ and solution qualities for problem instance $\mathcal{A}$. 
	Problem $\mathcal{A}$ consists of two items with weights $w_1=1$ and $w_2=1$ and knapsack capacity $W=1$. The values of the items are $c_1=1$ and $c_2=2$. The BKS is $101$ (marked in dark blue). The green filled circles correspond to the parameters sampled by the classical optimizer. The red cross marks the global minimum of the energy landscape, which is $-2.85$ and $-3.2$ for Figs. \ref{fig:landscape_bad} and \ref{fig:landscape_good}, respectively.}
	\label{fig:solution-landscapes}
\end{figure*}

In our approach we use an adapted CE method to optimize the penalty values $A$ and $B$ of our problem Hamiltonian, Eq. \eqref{eq:objective_function_knapsack_ising_ex_total}. A variation of these values significantly changes the energy landscape of the optimization problem, and hence also influences the pathway of the optimizer. An example can be seen in Fig. \ref{fig:solution-landscapes}. In Fig. \ref{fig:landscape_bad} one can see comparatively bad selected penalty values while in Fig. \ref{fig:landscape_good} optimized ones, found by CE, were used. The histograms in Fig. \ref{fig:histogram_bad} and \ref{fig:histogram_good} show the corresponding solution qualities, w.r.t the best known solution (BKS). In Fig. \ref{fig:landscape_good} the classical optimizer is able to sample (marked with green circles) near the global minimum (marked with red cross), which reflects in an improved solution quality in Fig. \ref{fig:histogram_good}. Therefore, optimizing the penalty values of the problem Hamiltonian, makes it easier for the classical optimizer to find better gate parameters for the quantum circuit and consequently reach a better solution quality. 

\begin{figure}[ht]
\centerline{\includegraphics[width=1.0\columnwidth]{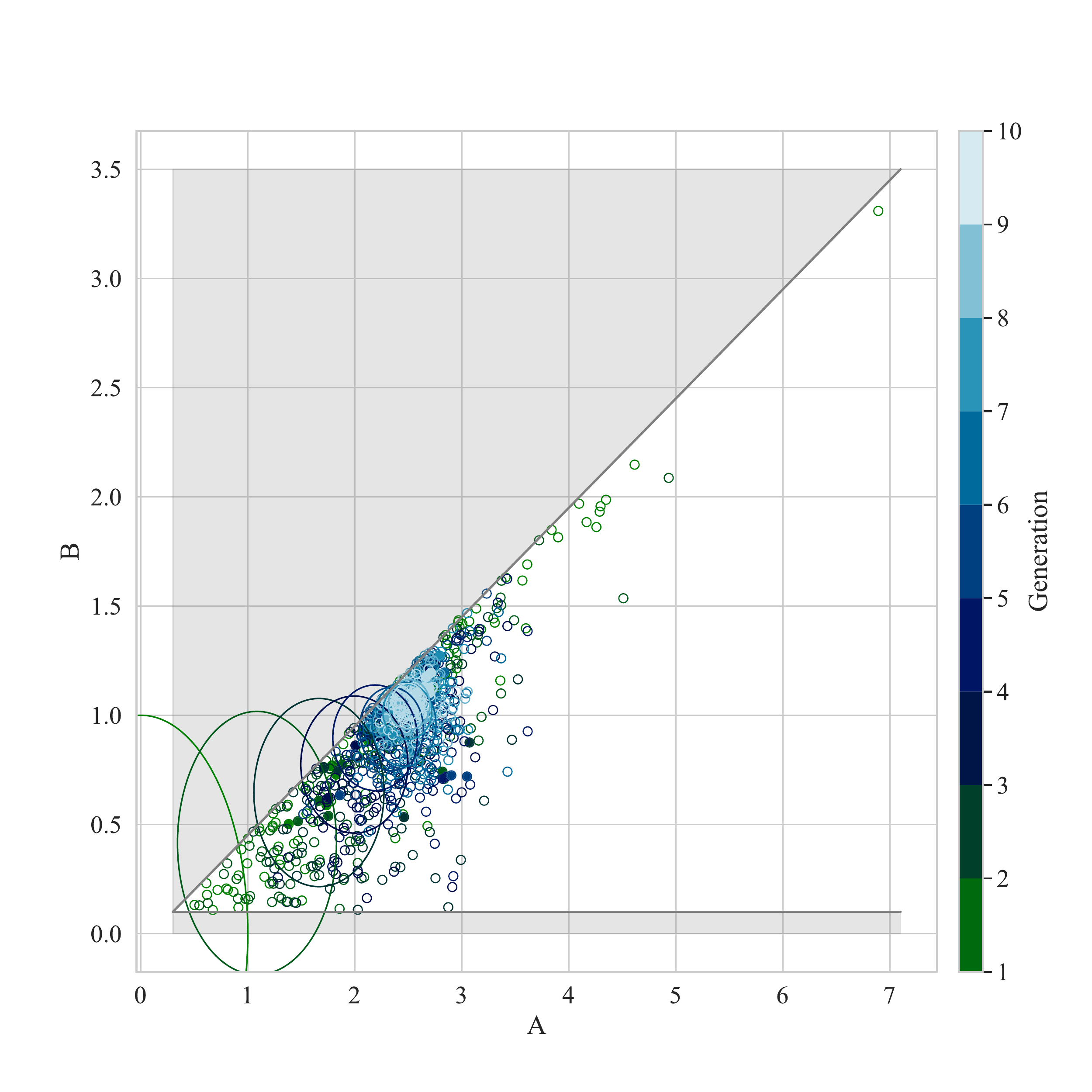}}
\caption{Example CE with $G=10$. The ellipses represent the $\mu_g$ and $\sigma_g^2$ of generation $g$. The filled circles correspond to the best $\rho$ fraction of individuals. The best values found by CE, for this specific problem instance $\mathcal{A}$, were $2.7$ and $1.1$ for $A$, respectively $B$.}
\label{fig:ellipse}
\end{figure}

Regarding the KP, the choice of penalty values is restricted by $0 < B \cdot \text{max}(c_\alpha) < A$, as stated in \cite{lucas2014ising}.
To satisfy this constraint, we use a modified CE optimization scheme (see Alg. \ref{algorithm:qaoa-cross-entropy}), in which the penalty values $A, B$ are sampled from truncated normal distributions $\mathpzc{p}$. Since the allowed values for $A$ depend on the choice of $B$, we first draw a value for $B$ (line 3) with an appropriately chosen sampling range $\Gamma_A$. Afterwards, the value for $A$ is drawn over a sampling range $\Gamma_A(B)$, such that the penalization constraint $0 < B \cdot \text{max}(c_\alpha) < A$ is satisfied (line 5). This is done for $n$ samples. For each sample, we construct the corresponding Hamiltonian, as described in Section \ref{subsec:kp-to-ising} and run the QAOA to assign a value $v_1...v_n$ corresponding to the approximation ratio of the best found solution for each $A_i, B_i$-pair (line 7). This is done iteratively for the specified number of generations $G$. In Fig. \ref{fig:ellipse}, the process of finding optimal penalty values for the problem Hamiltonian with QAOA-CE is demonstrated. 

\begin{algorithm}[!ht]
	\caption{Cross-Entropy Penalty Optimization}\label{algorithm:qaoa-cross-entropy}
	\begin{algorithmic}
		\Function{OPTIMIZE}{$\mathpzc{p},\Phi_{0},f,\rho,n,G$}
		\For{$g = 1 \to G$}
		\State $B_1...B_n \sim \mathpzc{p}(\cdot|\Phi_{g-1}, \Gamma_B)$
		\State $\textbf{B}\gets B_1...B_n$
		\State $A_1...A_n \sim \mathpzc{p}(\cdot|\Phi_{g-1}, \Gamma_A(\textbf{B}))$
		\State $\textbf{A}\gets A_1...A_n$
		\State $v_1...v_n \sim QAOA(A_1, B_1)...QAOA(A_n, B_n)$ 
		\State $\textbf{v}\gets v_1...v_n$
		\State $\text{sort \textbf{A, B} according to \textbf{v}}$
		\State $\Phi_{g} \gets \text{argmax}_{\Phi}\prod_{i=1}^{\lceil n\rho\rceil} \mathpzc{p}(A_i,B_i|\Phi)$
		\EndFor
		\Return $A_1, B_1$
		\EndFunction
	\end{algorithmic}
\end{algorithm}

\section{Evaluation}
\subsection{Experimental Setup}

For the experimental evaluation, we used the high-performance simulator Qiskit Aer \cite{aleksandrowicz2019qiskit} to perform simulations of QAOA circuits. Qiskit also wraps different optimization algorithms from the NLopt library \cite{johnson2014nlopt}, which can be used as the classical optimizer for the QAOA. We tried different ones, however, since they all performed quite similar on our KP instances, we chose the ESCH optimizer, which is a modified Evolutionary Algorithm for global optimization \cite{da2010designing}. 

In Tab. \ref{tab:ce-parameters} the parameter settings of the CE method are listed.
Since we are using a truncated normal distribution to sample from, we need to specify additional clipping parameters for both penalty values $A$ and $B$. The sampling range of $A$ is computed according to ${\Gamma_A(B) = \left[ B\cdot max(c_\alpha)+0.1, B\cdot max(c_\alpha)+10.0 \right]}$.   

\begin{table}[htbp]
\caption{Cross-Entropy Parameter Settings}
\begin{center}
\begin{tabular}{|c|c|c|c|c|c|c|c|}
\hline
\multicolumn{8}{|c|}{\textbf{CE Attributes}} \\
\cline{1-8} 
$G$ & $n$ & $\rho$ & $\gamma^{\mathrm{*}}$ & min $\sigma^2$ & $\sigma_0^2$ & $\mu_0$ & sampling range of B \\
\hline
10 & 100 & 0.1 & 0.5 & 0.1 & 1.0 & 0.0 & $\left[0.1,10.0\right]$ \\
\hline
\multicolumn{8}{l}{$^{\mathrm{*}}$The learning rate specifies the amount of changes from $\Phi_{g-1}$ to $\Phi_{g}$.}
\end{tabular}
\label{tab:ce-parameters}
\end{center}
\end{table}

We used five Knapsack Problem instances to test our QAOA-CE approach, which are stated in Tab. \ref{tab:datasets}. Since gate model hardware is still in its infancy and the conventional QAOA in general had a very low solution quality for larger Knapsack Problem instances, we rather used small instances, to demonstrate the effect of CE. However, our approach is theoretically also applicable to those larger problems. 

\begin{table}[ht]
\caption{Knapsack Problem Instances}
\begin{center}
\begin{tabular}{|c|c|c|c|c|c|}
\hline
\textbf{Problem}&\multicolumn{5}{|c|}{\textbf{Problem Attributes}} \\
\cline{2-6} 
\textbf{Instance} & \textbf{\textit{Items}} & \textbf{\textit{Weights}} & \textbf{\textit{Values}} & \textbf{\textit{W}} & \textbf{\textit{BKS}} \\
\hline
$\mathcal{A}$ & 2 & 1,1 & 2,1 & 1 & 101 \\
\hline
$\mathcal{B}$ & 2 & 1,1 & 1,2 & 1 & 011 \\
\hline
$\mathcal{C}$ & 2 & 1,1 & 2,1 & 2 & 1101 \\
\hline
$\mathcal{D}$ & 2 & 2,3 & 2,1 & 2 & 1001 \\
\hline
$\mathcal{E}$ & 2 & 1,2 & 2,1 & 2 & 1010 \\
\hline
\end{tabular}
\label{tab:datasets}
\end{center}
\end{table}

\subsection{Results \& Discussion}

\begin{figure*}[ht]
	\subfloat{\label{fig:sol-quality-p-1}
	\begin{overpic}[,width=1.0\textwidth]
    {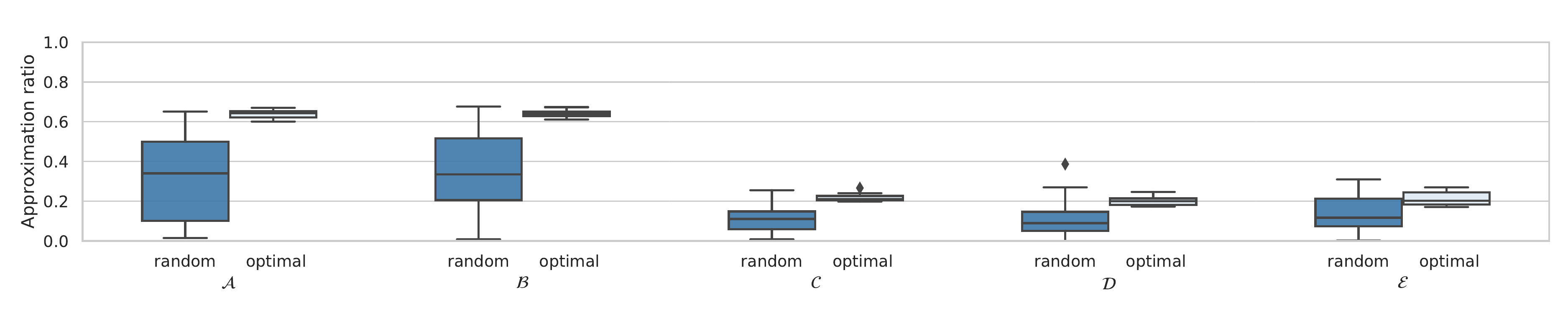}
     \put(1,20){(a)}
    \end{overpic}
  }
  \vspace{-1em}
    \\
    \subfloat{\label{fig:sol-quality-p-2}
	\begin{overpic}[,width=1.0\textwidth]
    {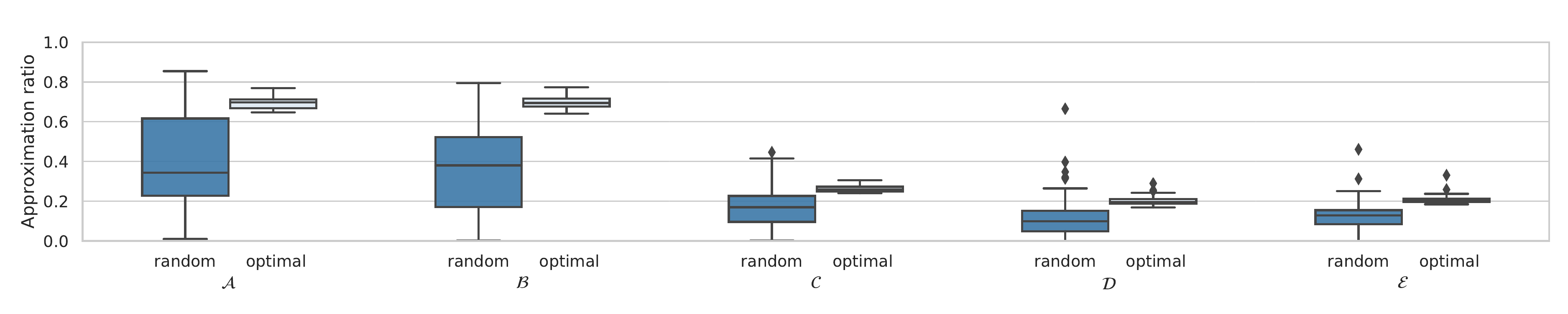}
     \put(1,20){(b)}
    \end{overpic}
 }
   \vspace{-1em}
    \\
    \subfloat{\label{fig:sol-quality-p-3}
	\begin{overpic}[,width=1.0\textwidth]
    {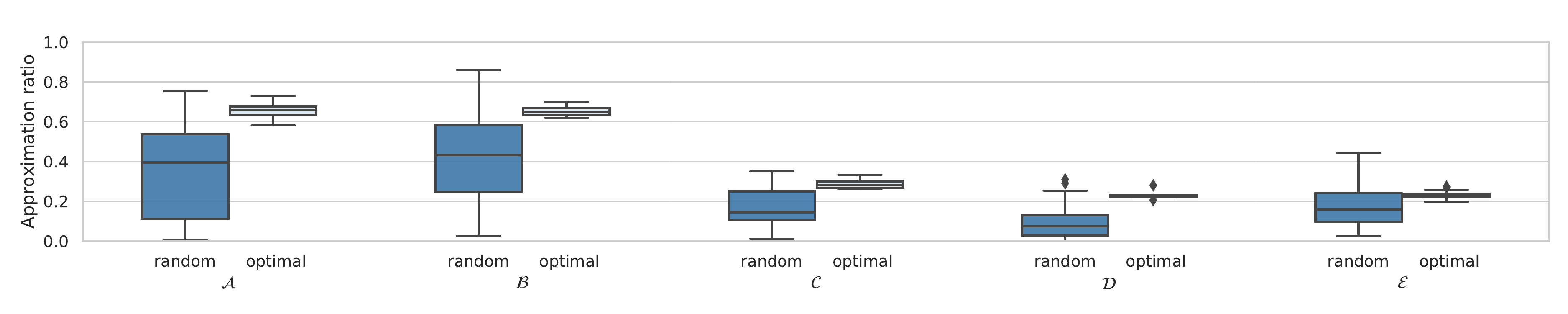}
     \put(1,20){(c)}
    \end{overpic}
  }
    \caption{Solution qualities for different QAOA circuit depth p=1,2,3 represented in \ref{fig:sol-quality-p-1}-\ref{fig:sol-quality-p-3}, respectively. For each circuit depth five KP instances $\mathcal{A}$-$\mathcal{E}$ were used. The ``random'' boxplots represent the approximation ratio of five randomly sampled penalty value pairs (each run 10 times), while the ``optimal'' boxplots represent the approximation ratio of the $\rho$ fraction of the penalty value pair population of the 10th generation of the CE method.}
    \label{fig:solution-qualities}
\end{figure*}

In Fig. \ref{fig:solution-qualities} the results of QAOA-CE and the conventional QAOA approach are compared against each other. We tested both methods on different problem instances (as stated in Tab. \ref{tab:datasets}). Fig. \ref{fig:sol-quality-p-1}-\ref{fig:sol-quality-p-3} differ in the QAOA 
circuit depth $p=1,2,3$. CE was initialized with the parameter setting of Tab. \ref{tab:ce-parameters} and the classical QAOA optimizer ESCH was set to 200 iterations. W.r.t the classical QAOA approach, we randomly sampled five penalty value pairs of the same sampling range as stated in Tab. \ref{tab:ce-parameters}. Each penalty value pair was executed 10 times and the corresponding solution qualities (approximation ratio) are represented in the respective box plot ``random''. For the QAOA-CE box plots, called ``optimal'', we used the $\rho$ fraction of the penalty value population of the last CE generation and calculated their fitness, i.e. solution quality. The solution quality is given by the approximation ratio of the BKS and can be calculated by dividing the BKS counts by the number of measurements (default 1024).

The results show, that for QAOA circuits with $p=1$, the solution quality, w.r.t the mean could be increased by more than $100\%$ in every problem instance by using the optimized penalty values (see Fig. \ref{fig:sol-quality-p-1}). Also for $p=2$ in Fig. \ref{fig:sol-quality-p-2}, the solution qualities were increased by around $100\%$. In Fig. \ref{fig:sol-quality-p-3} a significant growth in the quality of the solution can be seen, too, for $p=3$. However, since the size of the solution landscape of gate parameters increases polynomially with the circuit depth $p$, it is more difficult for the classical method to find the optimal parameters. That is why the solution quality improvement is  less significant in comparison.

Another aspect, which can be seen in Fig. \ref{fig:solution-qualities}, is the comparatively small variance of the approximation ratio of the optimized penalty values, throughout all problem instances. That is because those values are picked from the best $\rho$ fraction of individuals of the last CE generation, while the five random sampled penalty value pairs may contain unfavorable pairs. Nevertheless this does not detract from the fact, that our CE method is quite stable regarding the stochastic classical optimizer.

Additionally we have to mention that with using randomized penalty values there is sometimes the opportunity to outperform the optimized quite stable penalty values (see for example the maximum approximation ratio in Fig. \ref{fig:sol-quality-p-3} for problem instance $\mathcal{B}$ or $\mathcal{E}$). However this is due to the very stochastic system.

Since the penalty values found by the CE method differ in the used problem instances, it is hard to give a generally valid statement about inherent correlations. Hence the penalty values have to be optimized for each problem instance individually. As a result of this, we see the strength of our approach in the optimization of the overall solution qualities, but less in achieving computational speedups.



\section{Conclusion}
In this work we proposed a Cross Entropy approach for optimizing the penalty values of a Knapsack problem Hamiltonian. This allowed the classical optimizer of the QAOA to find better variational parameters for the quantum circuit and hence reach a significantly better solution quality compared to not using CE. We were able to show that for circuit depths $p=1$ and $p=2$ there is an enhancement in the solution quality of more than $100\%$. Additionally, choosing the optimized penalty values of the last CE generation results in a smaller variance of the solution quality, which implies that fewer averages have to be taken in order to find good solutions.
Since we did not come across a correlation between the optimized penalty values for different problem instances, it would be interesting to see if certain pattern recognition algorithms are able to find possible correlations. Additionally, we want to investigate our approach also for larger problem instances, when quantum hardware gets bigger and is capable of solving such instances.
Furthermore, it would be worth applying QAOA-CE also to other constrained optimization problems, like e.g. the Traveling Salesman Problem or the Set Cover Problem. 

\bibliographystyle{IEEEtran} 
\bibliography{IEEEbibliography}

\begin{thebibliography}{10}
\providecommand{\url}[1]{#1}
\csname url@samestyle\endcsname
\providecommand{\newblock}{\relax}
\providecommand{\bibinfo}[2]{#2}
\providecommand{\BIBentrySTDinterwordspacing}{\spaceskip=0pt\relax}
\providecommand{\BIBentryALTinterwordstretchfactor}{4}
\providecommand{\BIBentryALTinterwordspacing}{\spaceskip=\fontdimen2\font plus
\BIBentryALTinterwordstretchfactor\fontdimen3\font minus
  \fontdimen4\font\relax}
\providecommand{\BIBforeignlanguage}[2]{{%
\expandafter\ifx\csname l@#1\endcsname\relax
\typeout{** WARNING: IEEEtran.bst: No hyphenation pattern has been}%
\typeout{** loaded for the language `#1'. Using the pattern for}%
\typeout{** the default language instead.}%
\else
\language=\csname l@#1\endcsname
\fi
#2}}
\providecommand{\BIBdecl}{\relax}
\BIBdecl

\bibitem{IBMLaunch}
``Ibm makes quantum computing available on ibm cloud to accelerate
  innovation,''
  \url{https://www-03.ibm.com/press/us/en/pressrelease/49661.wss}, Accessed:
  2019-01-30.

\bibitem{DwaveLaunch}
``D-wave systems sells its first quantum computing system to lockheed martin
  corporation,''
  \url{https://www.dwavesys.com/news/d-wave-systems-sells-its-first-quantum-computing-system-lockheed-martin-corporation},
  Accessed: 2019-01-30.

\bibitem{grover1996fast}
L.~K. Grover, ``A fast quantum mechanical algorithm for database search,''
  \emph{arXiv preprint quant-ph/9605043}, 1996.

\bibitem{shor1999polynomial}
P.~W. Shor, ``Polynomial-time algorithms for prime factorization and discrete
  logarithms on a quantum computer,'' \emph{SIAM review}, vol.~41, no.~2, pp.
  303--332, 1999.

\bibitem{roetteler2017quantum}
M.~Roetteler, M.~Naehrig, K.~M. Svore, and K.~Lauter, ``Quantum resource
  estimates for computing elliptic curve discrete logarithms,'' in
  \emph{International Conference on the Theory and Application of Cryptology
  and Information Security}.\hskip 1em plus 0.5em minus 0.4em\relax Springer,
  2017, pp. 241--270.

\bibitem{peruzzo2014variational}
A.~Peruzzo, J.~McClean, P.~Shadbolt, M.-H. Yung, X.-Q. Zhou, P.~J. Love,
  A.~Aspuru-Guzik, and J.~L. O’brien, ``A variational eigenvalue solver on a
  photonic quantum processor,'' \emph{Nature communications}, vol.~5, p. 4213,
  2014.

\bibitem{farhi2014quantum}
E.~Farhi, J.~Goldstone, and S.~Gutmann, ``A quantum approximate optimization
  algorithm,'' \emph{arXiv preprint arXiv:1411.4028}, 2014.

\bibitem{zhou2018quantum}
L.~Zhou, S.-T. Wang, S.~Choi, H.~Pichler, and M.~D. Lukin, ``Quantum
  approximate optimization algorithm: performance, mechanism, and
  implementation on near-term devices,'' \emph{arXiv preprint
  arXiv:1812.01041}, 2018.

\bibitem{martello1990knapsack}
S.~Martello, ``Knapsack problems: algorithms and computer implementations,''
  \emph{Wiley-Interscience series in discrete mathematics and optimiza tion},
  1990.

\bibitem{karp1972reducibility}
R.~M. Karp, ``Reducibility among combinatorial problems,'' in \emph{Complexity
  of computer computations}.\hskip 1em plus 0.5em minus 0.4em\relax Springer,
  1972, pp. 85--103.

\bibitem{mcgeoch2014adiabatic}
C.~C. McGeoch, ``Adiabatic quantum computation and quantum annealing: Theory
  and practice,'' \emph{Synthesis Lectures on Quantum Computing}, vol.~5,
  no.~2, pp. 1--93, 2014.

\bibitem{lucas2014ising}
A.~Lucas, ``Ising formulations of many np problems,'' \emph{Frontiers in
  Physics}, vol.~2, p.~5, 2014.

\bibitem{suzuki1976generalized}
M.~Suzuki, ``Generalized trotter's formula and systematic approximants of
  exponential operators and inner derivations with applications to many-body
  problems,'' \emph{Communications in Mathematical Physics}, vol.~51, no.~2,
  pp. 183--190, 1976.

\bibitem{shaydulin2019multistart}
R.~Shaydulin, I.~Safro, and J.~Larson, ``Multistart methods for quantum
  approximate optimization,'' in \emph{2019 IEEE High Performance Extreme
  Computing Conference (HPEC)}.\hskip 1em plus 0.5em minus 0.4em\relax IEEE,
  2019, pp. 1--8.

\bibitem{rubinstein2004cross}
R.~Y. Rubinstein and D.~P. Kroese, ``The cross-entropy method: A unified
  approach to monte carlo simulation, randomized optimization and machine
  learning,'' \emph{Information Science \& Statistics, Springer Verlag, NY},
  2004.

\bibitem{rubinstein1999cross}
R.~Rubinstein, ``The cross-entropy method for combinatorial and continuous
  optimization,'' \emph{Methodology and computing in applied probability},
  vol.~1, no.~2, pp. 127--190, 1999.

\bibitem{weinstein2013open}
A.~Weinstein and M.~L. Littman, ``Open-loop planning in large-scale stochastic
  domains,'' in \emph{Twenty-Seventh AAAI Conference on Artificial
  Intelligence}, 2013.

\bibitem{barak2015beating}
B.~Barak, A.~Moitra, R.~O'Donnell, P.~Raghavendra, O.~Regev, D.~Steurer,
  L.~Trevisan, A.~Vijayaraghavan, D.~Witmer, and J.~Wright, ``Beating the
  random assignment on constraint satisfaction problems of bounded degree,''
  \emph{arXiv preprint arXiv:1505.03424}, 2015.

\bibitem{crooks2018performance}
G.~E. Crooks, ``Performance of the quantum approximate optimization algorithm
  on the maximum cut problem,'' \emph{arXiv preprint arXiv:1811.08419}, 2018.

\bibitem{goemans1995improved}
M.~X. Goemans and D.~P. Williamson, ``Improved approximation algorithms for
  maximum cut and satisfiability problems using semidefinite programming,''
  \emph{Journal of the ACM (JACM)}, vol.~42, no.~6, pp. 1115--1145, 1995.

\bibitem{guerreschi2019qaoa}
G.~G. Guerreschi and A.~Matsuura, ``Qaoa for max-cut requires hundreds of
  qubits for quantum speed-up,'' \emph{Scientific reports}, vol.~9, no.~1, p.
  6903, 2019.

\bibitem{preskill2018quantum}
J.~Preskill, ``Quantum computing in the nisq era and beyond,'' \emph{Quantum},
  vol.~2, p.~79, 2018.

\bibitem{hadfield2019quantum}
S.~Hadfield, Z.~Wang, B.~O'Gorman, E.~G. Rieffel, D.~Venturelli, and R.~Biswas,
  ``From the quantum approximate optimization algorithm to a quantum
  alternating operator ansatz,'' \emph{Algorithms}, vol.~12, no.~2, p.~34,
  2019.

\bibitem{de2019knapsack}
P.~D. de~la Grand'rive and J.-F. Hullo, ``Knapsack problem variants of qaoa for
  battery revenue optimisation,'' \emph{arXiv preprint arXiv:1908.02210}, 2019.

\bibitem{aleksandrowicz2019qiskit}
G.~Aleksandrowicz, T.~Alexander, P.~Barkoutsos, L.~Bello, Y.~Ben-Haim,
  D.~Bucher, F.~Cabrera-Hern{\'a}ndez, J.~Carballo-Franquis, A.~Chen, C.~Chen
  \emph{et~al.}, ``Qiskit: An open-source framework for quantum computing,''
  \emph{Accessed on: Mar}, vol.~16, 2019.

\bibitem{johnson2014nlopt}
S.~G. Johnson, ``The nlopt nonlinear-optimization package,'' 2014.

\bibitem{da2010designing}
C.~H. da~Silva~Santos, M.~S. Goncalves, and H.~E. Hernandez-Figueroa,
  ``Designing novel photonic devices by bio-inspired computing,'' \emph{IEEE
  Photonics Technology Letters}, vol.~22, no.~15, pp. 1177--1179, 2010.

\end{thebibliography}

\vspace{12pt}
\color{red}

\end{document}